\pdfoutput=1    

\documentclass[prl,twocolumn,showpacs,preprintnumbers,amsmath,amssymb,superscriptaddress]{revtex4}

\usepackage[utf8]{inputenc}  
\usepackage{graphicx}        
\usepackage{dcolumn}         
\usepackage{bm}              
\usepackage{xcolor}          

\graphicspath{{Figures/}}         

\newcommand{\mbold}[1]{\mbox{\boldmath${#1}$}}
\renewcommand{\vec}[1]{\mathbf{{#1}}}
\newcommand{\pvec}[1]{\mathbf{{#1}}_\parallel }

\begin{document}


\title{The Scattering of Electromagnetic Waves from Two-Dimensional
  Randomly Rough Penetrable Surfaces}

\author{Ingve Simonsen}
\email{Ingve.Simonsen@ntnu.no}
\affiliation{Department of Physics, 
             Norwegian University of Science and Technology (NTNU), 
             NO-7491 Trondheim, Norway}

\author{Alexei A. Maradudin} 

\author{Tamara A. Leskova} 

\affiliation{Department of Physics and Astronomy and Institute for
  Surface and Interface Science, University of California, Irvine CA  92697, U.S.A.}

\date{\today}

\begin{abstract}
  An accurate and efficient numerical simulation approach to
  electromagnetic wave scattering from two-dimensional, randomly
  rough, penetrable surfaces is presented.  The use of the M\"uller
  equations and an impedance boundary condition for a two-dimensional
  rough surface yields a pair of coupled two-dimensional integral
  equations for the sources on the surface in terms of which the
  scattered field is expressed through the Franz formulas.  By this
  approach, we calculate the full angular intensity distribution of
  the scattered field that is due to a finite incident beam of
  $p$-polarized light. We specifically check the energy conservation
  (unitarity) of our simulations (for the non-absorbing case). Only
  after a detailed numerical treatment of {\em both} diagonal and
  close-to-diagonal matrix elements is the unitarity condition found
  to be well-satisfied for the non-absorbing case (${\mathcal
    U}>0.995$), a result that testifies to the accuracy of our
  approach.
\end{abstract}

\maketitle


The scattering of electromagnetic waves from two-dimensional randomly
rough penetrable surfaces has been studied theoretically for more than
five decades. The methods used in these studies in recent years, where
attention has been directed toward multiple-scattering phenomena, have
been either analytical in nature or computational.  Chief among the
former methods has been the small-amplitude perturbation
theory~\cite{1,2,3}, while several different computational methods
have been used in solving the scattering problem.  In the earliest
calculation of this type~\cite{4} the system of coupled inhomogeneous
integral equations for the tangential components of the total electric
and magnetic fields on the rough surface obtained from scattering
theory, was converted into a system of inhomogeneous matrix equations
by the use of the method of moments~\cite{5}, which was then solved by
Neumann-Liouville iteration~\cite{6}.
This is a formally exact approach but one that is computationally (and
memory) intensive.

In subsequent work on this problem approximate solutions of the exact
integral equations have been sought. In the sparse-matrix flat-surface
iterative approach~\cite{7,8} the matrix elements for two close points
on the surface are treated exactly, while those connecting two widely
separated points are treated approximately, in an iterative solution
of the matrix equations.
 
Soriano and Saillard~\cite{9} have combined the sparse-matrix
flat-surface iterative approach with an impedance
approximation~\cite{10} to calculate co-polarized and cross-polarized
bistatic scattering coefficients of aluminum randomly rough surfaces
for comparison with results obtained from perfectly conducting
surfaces.
 
An approach that combines the fast multipole method~\cite{11} and the
sparse-matrix flat-surface iterative approach has been developed by
Jandhyala {\it et. al}~\cite{12}. 
  
Despite these advances, the calculation of the scattering of
electromagnetic waves from two-dimensional, penetrable, randomly rough
surfaces, remains a computationally intensive problem, in need of
further improvements in the methods used to solve it.

In this paper we use the Franz formulas of electromagnetic scattering
theory~\cite{13,16} to obtain expressions for the amplitude of the
electromagnetic field scattered from a two-dimensional, rough,
metallic or dielectric surface in terms of the tangential components
of the total electric and magnetic fields on the surface. The
independent elements of these tangential field components satisfy a
system of four coupled inhomogeneous two-dimensional integral
equations --- the M\"uller equations~\cite{14,15} --- derived from
Franz formulas.  This system of four integral equations is reduced to
a system of two integral equations by the use of an impedance boundary
condition at a two-dimensional rough metallic surface~\cite{17}, and
its solution is used to calculate the mean differential reflection
coefficient.

\begin{figure}[tbh]
  \begin{center}
    \includegraphics*[width=0.95\columnwidth,height=0.60\columnwidth]{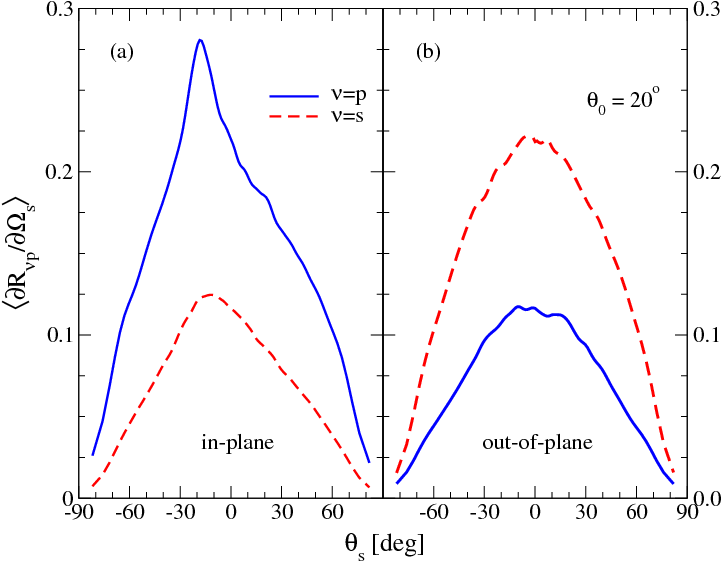}
  \end{center}
  \caption{(Color online) The mean differential reflection
    coefficients~\cite{20}, $\left<\partial R_{\nu
        p}/\partial\Omega_s\right>$ $(p \rightarrow \nu$) for a
    $p$-polarized incident beam, whose polar angle
    of incidence is $\theta_0=20^\circ$, as functions of the polar
    scattering angle $\theta_s$ for the (a) in-plane and (b)
    out-of-plane scattering.  See text for additional parameters.}
  \label{Fig:1}
\end{figure}

The approach to the scattering of an electromagnetic field from a
rough metallic or dielectric surface outlined here is similar to the
approach of Soriano and Saillard~\cite{5} in its use of an impedance
boundary condition to reduce the number of coupled integral equations
that have to be solved from four to two. However, there are still
important differences between these two approaches. The first is that
we do not use the sparse-matrix flat-surface iterative approach: the
matrix elements connecting any two points on the surface are
calculated accurately. Moreover, those connecting two nearby points
are calculated more accurately than in the work of Soriano and
Saillard.  The second is that we calculate the full angular intensity
distribution of the scattered field, which allows us to check the
unitarity (energy conservation) of the scattered field (for the
non-absorbing case). This enabled the identification of critical
aspects of the numerical implementation that, if not handled properly,
could lead to erroneous results and/or a significant drop in
accuracy. This important point seems to have been overlooked in
previous publications. The third is that although the occurrence of
hyper-singular kernels is avoided in both approaches by the use of the
M\"uller equations~\cite{14,15}, some differences are found between
our resulting matrix elements and those of Soriano and Saillard that
appear to affect the unitarity of the scattered
field\cite{24}. Moreover, contrary to what was reported in
Ref.~\cite{14}, we find that matrix element terms containing the
Green's function of the metal also have to be taken into account for
some off-diagonal elements in order to produce accurate results. The
fourth is that we do not use the beam decomposition method~\cite{19}
for the incident beam in which a wide beam is represented by a
weighted sum of shifted narrow beams. Instead we use a single wide
incident beam.

%
%
The physical system that we consider consists of vacuum
[$\varepsilon_0=1$] in the region $x_3>\zeta(\pvec{x})$ [where
$\pvec{x}=(x_1,x_2,0)$] and a non-magnetic metal in the region
$x_3<\zeta(\pvec{x})$ that is characterized by a complex,
frequency-dependent, dielectric function, $\varepsilon(\omega)$, for
which $Re\,\varepsilon(\omega)<0$ and $Im\,\varepsilon(\omega)>0$.
The surface profile function, $x_3=\zeta(\pvec{x})$, is assumed to
constitute a zero-mean, Gaussian random process that is a
single-valued function of $\pvec{x}$ and is differentiable with
respect to $x_1$ and $x_2$ at least twice. It is defined by $\left<
  \zeta(\pvec{x})\right>=0$ and $\left< \zeta(\pvec{x})
  \zeta(\pvec{x}') \right>=\delta^2W(|\pvec{x}-\pvec{x}'|)$, where
$\delta$ is the root-mean-square roughness, $W(\cdot)$ denotes the
(normalized) correlation function, and the angle brackets denote an
average over the ensemble of realizations of $\zeta(\pvec{x})$. In
this work we will use an isotropic Gaussian correlation function
$W(x)=\exp(-x^2/a^2)$ with $a$ the correlation length.

%
%
The starting point of our approach is the Franz formulas of
electromagnetic theory (or the dyadic form of Huygens
principle)~\cite{13,16}. By applying them to the vacuum region above
the metal surface, and letting the observation point, $\vec{x}$,
approach the surface $x_3=\zeta(\pvec{x})$, two coupled inhomogeneous
integral equations for the tangential components of the electric and
magnetic fields,
$\vec{J}_E(\pvec{x})=\left.\hat{\vec{n}}\times\vec{E}(\vec{x})\right|_{x_3=\zeta(\pvec{x})}$
and
$\vec{J}_H(\pvec{x})=\left.\hat{\vec{n}}\times\vec{H}(\vec{x})\right|_{x_3=\zeta(\pvec{x})}$,
respectively, are obtained, where $\hat{\vec{n}}$ denotes the unit
vector normal to the surface and directed into the vacuum. These
equations contain double derivatives of the scalar Green's function
$g_0(\vec{x},\vec{x}')=\exp[ik_0R]/4\pi R$
[$k_0=\sqrt{\varepsilon_0(\omega)}\,\omega/c$, $R=|\vec{x}-\vec{x}'|$]
resulting in non-integrable {\em hyper-singular} kernels that are
sources of computational difficulties~\cite{9,14,15}. One way to
obtain integrable kernels, is to combine in a suitable way the two
sets of Franz formulas obtained separately for the vacuum and metal
regions so that the resulting integral equations do not contain {\em
  any} hyper-singular terms.  The resulting equations are known as the
M\"uller integral equations~\cite{14}, and the one satisfied by
$\vec{J}_H(\pvec{x}|\omega)$ reads
\begin{align}
  \label{eq:Muller}
  & \vec{J}_H(\pvec{x}|\omega) 
  =
  \vec{J}_H(\pvec{x}|\omega)_{inc}
  \nonumber \\ & \;  
 +{\mathcal P} \int \! d^2x_\parallel' \,
  [\![     
    \hat{\vec{n}}(\pvec{x}) \times 
       \{  
       \mbold{\nabla}\times
       [   
          {\mathcal G}(\vec{x}| \vec{x}') \, 
          \vec{J}_H(\pvec{x}'|\omega) 
       ]   
      \}   
  ]\!]     
  \\ & \; 
 - \frac{ic}{\omega}
  \int \! d^2x_\parallel' \,
  [\![ 
    \hat{\vec{n}}(\pvec{x}) \times 
     \{  
       \mbold{\nabla} \times \mbold{\nabla}\times
      [  
         {\mathcal G}(\vec{x}| \vec{x}')\,
          \vec{J}_E(\pvec{x}'|\omega) 
     ]   
     \}  
     ]\!], 
   \nonumber
\end{align}
where the equation satisfied by $\vec{J}_E(\pvec{x}'|\omega)$ can be
obtained from duality~\cite{16}. In writing
Eq.~(\ref{eq:Muller}) we have introduced ${\mathcal G}(\pvec{x}|
\pvec{x}')= g_0(\pvec{x}| \pvec{x}') - g(\pvec{x}| \pvec{x}')$ --- the
difference between the scalar Green's functions for the vacuum
(subscript~$0$) and the metal; $[\![ A(\vec{x}| \vec{x}') ]\!] =
\left. A(\vec{x}| \vec{x}')\right|_{x_3=\zeta(\pvec{x});
  x_3'=\zeta(\pvec{x}')}$; ${\mathcal P}$ denotes the Cauchy principle
value of an integral; and $\vec{J}_H(\pvec{x}|\omega)_{inc}$ is
defined similarly to $\vec{J}_H(\pvec{x}|\omega)$ but for the
incident field.  Initially the kernel of Eq.~(\ref{eq:Muller})
seems to be hyper-singular. However, because the leading term (when
$R\rightarrow 0$) of the second derivative of the
scalar Green's function is independent of medium parameters, the most divergent terms of
the kernel cancel, rendering it integrable.
%
By adopting the impedance boundary condition that relates the surface
currents $\vec{J}_E(\pvec{x}|\omega)$ and $\vec{J}_H(\pvec{x}|\omega)$
via the (local) impedance tensor (${\mathbf K}$)~\cite{17}:
$\vec{J}_E(\pvec{x}|\omega)_i = K_{ij}(\pvec{x}|\omega)
\vec{J}_H(\pvec{x}|\omega)_j$ [$i,j=1,2$], the dependence on
$\vec{J}_E(\pvec{x}|\omega)$ can be removed from
Eq.~(\ref{eq:Muller}). Moreover, the resulting equation can be
converted into a matrix equation for the two independent electric
surface current components, say, $\vec{J}_H(\pvec{x}|\omega)_i$
[$i=1,2$], by the use of the method of moments~\cite{5}. The resulting
linear system is then solved by the BiCGStab method~\cite{18} and the
solution used to calculate the mean differential reflection
coefficient that is averaged over an ensemble of realizations of the
surface profile function (see Ref.~\cite{20} for details).

%
%
On the basis of the integral
equation~(\ref{eq:Muller}), and with the use of the
impedance boundary condition, we have performed numerical simulations
for a $p$-polarized incident beam of wavelength
$\lambda=0.6328~\mu\mbox{m}$ that is scattered from a Gaussian
randomly rough silver surface. At this wavelength
$\varepsilon(\omega)=-16.00+i1.088$~\cite{21}.  The surface was
characterized by a root-mean-square roughness of $\delta=\lambda/4$
and a correlation length $a=\lambda/2$. In the simulations it was
assumed to cover an area of $16\lambda\times
16\lambda$, and the discretization interval was
$\Delta=\lambda/7$ for both the $x_1$- and
$x_2$-directions.

\begin{figure}[tbh]
  \begin{center}
    \includegraphics*[width=0.80\columnwidth]{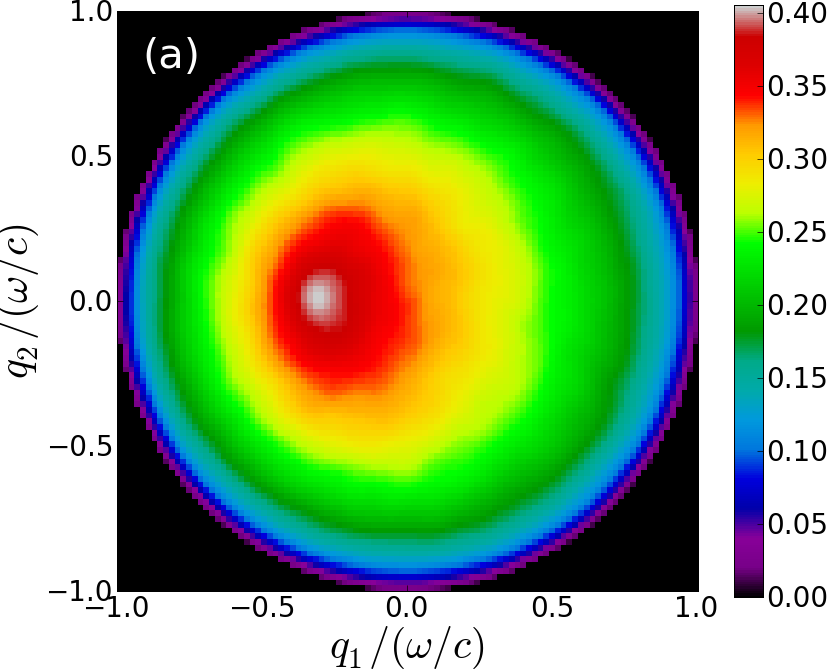} \\
    \includegraphics*[width=0.80\columnwidth]{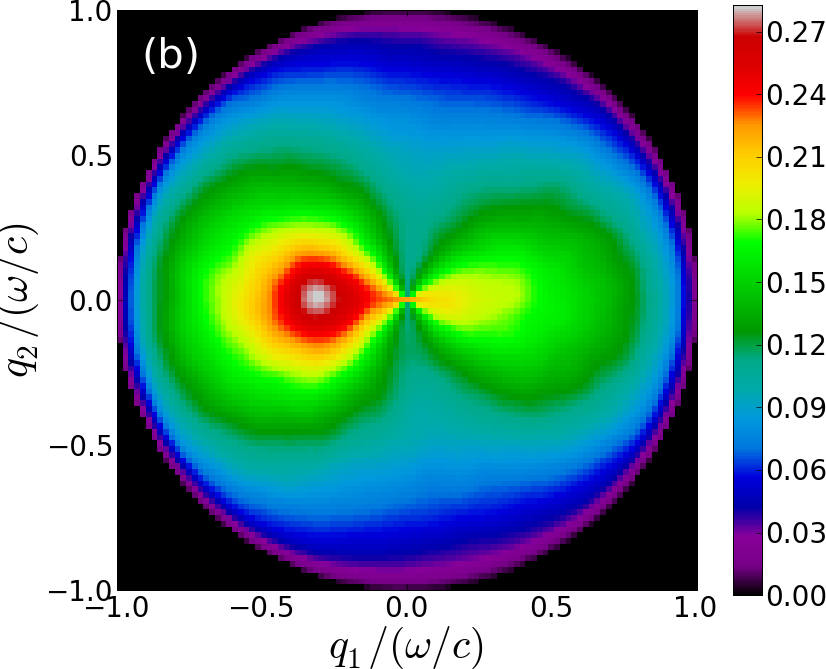}  \\
    \includegraphics*[width=0.80\columnwidth]{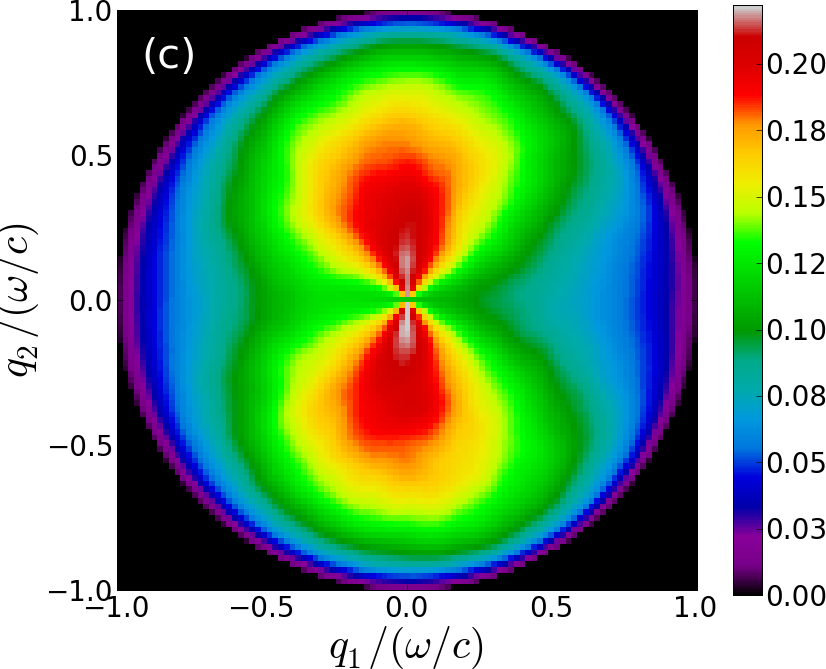}  
  \end{center}
  \caption{(Color online) 
    The same as Fig.~\protect\ref{Fig:1}, but now showing the full
    angular intensity distribution of the scattered field. (a)
    $p\rightarrow \mbox{polarization not recorded}$; (b) $p\rightarrow
    p$; and (c) $p\rightarrow s$.}  
  \label{Fig:2}
\end{figure}

Figure~\ref{Fig:1} presents the mean differential reflection
coefficients as functions of the polar scattering angle $\theta_s$ for
the in-plane~[Fig.~\ref{Fig:1}(a)] 
and out-of-plane ($\phi_s=\pm 90^{\circ}$)~[Fig.~\ref{Fig:1}(b)]
scattered light due to a $p$-polarized Gaussian beam of width
$w=4\lambda$ incident on the surface at a polar angle
$\theta_0=20^\circ$. For the same parameters, Figs.~\ref{Fig:2} depict
the full angular intensity distribution of the incident $p$-polarized
light that is scattered into $p$- and $s$-polarized light
(polarization not recorded)~[Fig.~\ref{Fig:2}(a)]; $p$-polarization
[Fig.~\ref{Fig:2}(b)]; and $s$-polarization [Fig.~\ref{Fig:2}(c)].
The number of surface realizations used to obtain these results was
$N_\zeta=5000$. The simulations required for every surface realization
$96$ CPU seconds (on a $2.67\mbox{GHz}$ Intel i7 CPU) for each angle
of incidence when calculating the scattered field on a $100\times100$
grid. The peaks observed in Figs.~\ref{Fig:1} and \ref{Fig:2} at
angular positions $\theta_s=-\theta_0$ (and $\phi_s=\phi_0+\pi$) are
due to the enhanced backscattering phenomenon, a multiple scattering
effect~\cite{22}.  The energy fraction of the incident light that is
scattered by the surface was $94.7\%$, compared to $96.9\%$ as
predicted from the Fresnel coefficient of the corresponding flat
interface scattering system. All the light scattered by the surface
was essentially incoherent (diffuse) (about $99.98\%$).

%
In order to evaluate the accuracy of the simulations and to perform a
self-consistency check of our approach, we have performed simulations
using the parameters given above with the exception that the metal was
assumed to be non-absorbing, {\it i.e.}  we artificially put
$\mbox{Im}\,\varepsilon(\omega)\equiv 0$.  Under this assumption, the total
time-averaged power fluxes of the incident and scattered fields have to
be equal, or in other word, one should require {\em energy
  conservation} (or equivalently unitarity of the scattered field,
${\mathcal U}\equiv 1$ where ${\mathcal U}$ denotes the fraction of
the incident power flux that is scatted by the rough metal surface. It
should be stressed that energy conservation is only a necessary, but
not sufficient condition to guarantee the correctness of the
simulation results for non-absorbing systems. It is still, however, a
rather useful and non-trivial condition that can assist in detecting
inaccuracies of the simulation approach as well as potential
implementation errors.  For the parameters used in the simulations
reported in this work, we found ${\mathcal U}>0.995$ for
``non-absorbing'' silver [$\varepsilon(\omega)=-16.00$], a result that
testifies to the accuracy of our approach.

In order to achieve such good unitarity values, it turned out that
great care had to be exercised when handling the matrix elements of
the integral equation kernel that were on, or close to, the
diagonal. Soriano and Saillard~\cite{9} have previously pointed out
one way of handling the diagonal matrix elements that contain the
singularity (at $\pvec{x}=\pvec{x}'$) of the Green's function by
separating it into two parts: one for which the integrand is singular
but integrable and is handled analytically, and another that is
regular and is handled numerically.  These authors were not able
within their approach to check the energy conservation of their
calculations. We have found, however, that in order to achieve good
results for the energy conservation, also {\em close-to} diagonal
matrix elements (in addition to those on the diagonal) have to be
treated accurately even if the off-diagonal matrix elements are
regular everywhere. These findings somewhat resemble results reported
for volume integral equations where also close to diagonal
volume-elements had to be handled with higher accuracy then more
distant matrix elements\cite{23}.  For instance the use of the
midpoint method for evaluating all off-diagonal matrix elements (as in
Ref.~\cite{9}) and a more accurate method for the diagonal elements,
would, for the surface parameters assumed here, result in about
$16.4\%$ (${\mathcal U}=0.834$) of the incident energy not being
accounted for, a result that was found to be more-or-less independent
of how accurately one treated the diagonal elements, or if the surface
was rough or flat. Moreover, if in addition to the diagonal matrix
elements, also the nearest-neighbor elements were treated accurately,
the amount of energy that was not accounted for dropped to $4.9\%$
(${\mathcal U}=0.951$). If, additionally, also next-nearest neighbor
matrix elements were treated accurately, the unitarity condition
started to become well satisfied (${\mathcal U}>0.995$), and treating
accurately matrix elements that were even further apart contributed
only insignificantly to the improvement of the unitarity
condition~\cite{24}.  It should be stressed that without performing
the self-consistency check based on energy conservation, which
requires the full angular distribution of the scattered light being
available to us, it has probably not been realized that failing to
treat close-to-diagonal matrix elements accurately could cause
inaccuracies in the range of $10$--$20\%$ even for flat
interfaces. This is one of the main results of this Letter.

%
%
\smallskip
In conclusion, we have presented an accurate and high-performance
simulation approach for the scattering of electromagnetic waves from
two-dimensional penetrable metallic surfaces based on surface integral
techniques. By this approach, the scattering of a $p$-polarized finite
beam by a two-dimensional, randomly rough, silver surface was studied
in the optical regime, and it gave rise to the well-known enhanced
backscattering phenomenon. Due to the calculation of the full angular
intensity distribution of the scattered light, it was possible for us
to evaluate the accuracy of the simulation approach.  It was found
that high-quality simulation results required an accurate treatment of
{\em both} the diagonal and close-to-diagonal matrix elements. This
latter point seems to have been overlooked in previous studies. In
this way, we were able to obtain results that respect energy
conservation (unitarity) for the equivalent non-absorbing system,
something that testifies to the accuracy of our approach.
                            
%
%
The simulation approach presented in this Letter opens the door to a
direct and detailed comparison of the full angle-resolved intensity
distributions of the scattered light obtained experimentally and
theoretically. Additionally, the approach provides the tools needed to
predict the effect of surface roughness on the electromagnetic field
in the near and far zone of the surface, and also to tailor surface
structures towards specific applications (engineered surfaces).  Such
issues are of importance in numerous applications, like for instance,
in the photovoltaic industry where surface roughness in solar cells is
known to increase the efficiency of the cell, but the optimal surface
structure, and the mechanism responsible for the increased efficiency,
are still unknown~\cite{25}.

%
%
\begin{acknowledgements}
  This research was supported in part by AFRL contract
  FA9453-08-C-0230, the Research Council of Norway (Sm{\aa}forsk
  grant), and NTNU.
\end{acknowledgements}


\end{document}